\documentclass[conference]{IEEEtran}
\IEEEoverridecommandlockouts
\usepackage{cite}
\usepackage{amsmath,amssymb,amsfonts}
\usepackage{algorithmic}
\usepackage{graphicx}
\usepackage{textcomp}
\usepackage{xcolor}
\usepackage{booktabs}
\usepackage{multirow}
\usepackage{tabularx}
\def\BibTeX{{\rm B\kern-.05em{\sc i\kern-.025em b}\kern-.08em
    T\kern-.1667em\lower.7ex\hbox{E}\kern-.125emX}}
\begin{document}

\title{\textsc{CorDEL}: A Contrastive Deep Learning Approach for Entity Linkage}

\author{\IEEEauthorblockN{Zhengyang Wang}
\IEEEauthorblockA{Texas A\&M University\\
College Station, TX, USA\\
zhengyang.wang@tamu.edu}
\and
\IEEEauthorblockN{Bunyamin Sisman, Hao Wei, Xin Luna Dong}
\IEEEauthorblockA{Amazon.com\\
Seattle, WA, USA\\
\{bunyamis, wehao, lunadong\}@amazon.com}
\and
\IEEEauthorblockN{Shuiwang Ji}
\IEEEauthorblockA{Texas A\&M University\\
College Station, TX, USA\\
sji@tamu.edu}
}

\maketitle

\begin{abstract}
    Entity linkage (EL) is a critical problem in data cleaning and integration. In the past several decades, EL has typically been done by rule-based systems or traditional machine learning models with hand-curated features, both of which heavily depend on manual human inputs. With the ever-increasing growth of new data, deep learning (DL) based approaches have been proposed to alleviate the high cost of EL associated with the traditional models. Existing exploration of DL models for EL strictly follows the well-known twin-network architecture. However, we argue that the twin-network architecture is sub-optimal to EL, leading to inherent drawbacks of existing models. In order to address the drawbacks, we propose a novel and generic contrastive DL framework for EL. The proposed framework is able to capture both syntactic and semantic matching signals and pays attention to subtle but critical differences. Based on the framework, we develop a contrastive DL approach for EL, called \textsc{CorDEL}, with three powerful variants. We evaluate \textsc{CorDEL} with extensive experiments conducted on both public benchmark datasets and a real-world dataset. \textsc{CorDEL} outperforms previous state-of-the-art models by 5.2$\%$ on public benchmark datasets. Moreover, \textsc{CorDEL} yields a 2.4$\%$ improvement over the current best DL model on the real-world dataset, while reducing the number of training parameters by 97.6$\%$. 
\end{abstract}


\section{Introduction}\label{sec:intro}

Entity linkage~(EL), also known as entity matching, record linkage, entity resolution, and duplicate detection, refers to the task of determining whether two data records represent the same real-world entity. For example, in a product database, a black ink tank for printers produced by Canon can be represented as (Black ink tank, Canon) with attributes (Product title, Brand). However, there exist many other ways to build records for the same product, such as (Ink tank [black], Canon) and (Black ink tank, Canon\textcircled{R} Ink). As a result, there might be many data records referring to the same real-world entity, needing to be cleaned and integrated. 

EL has been a fundamental problem in data cleaning and integration in many domains such as e-commerce~\cite{gokhale2014corleone,mudgal2018deep}, and data warehouses~\cite{winkler2009data}. Because of its importance, it has been extensively studied for several decades~\cite{dunn1946record,fellegi1969theory,elmagarmid2006duplicate,naumann2010introduction,christen2012data,getoor2012entity,sehili2015privacy,ebraheem2018distributed,mudgal2018deep,trivedi2018linknbed}. Models for EL have evolved with the development of machine learning~\cite{cohen2002learning,sarawagi2002interactive,bilenko2003adaptive,singla2006entity,konda2016magellan}, incorporating rule-based methods~\cite{wang2011entity,fan2009reasoning,singh2017synthesizing,singh2017generating} and crowd-sourcing~\cite{wang2012crowder,stonebraker2013data,vesdapunt2014crowdsourcing,gokhale2014corleone}.
However, because of the explosion in the volume and diversity of data, we are still far away from solving EL. Newly generated data may have different data distributions, requiring new models and thus a lot of human resources. For example, traditional machine learning models, such as support vector machines and random forests, usually require humans to hand-craft features for different data to maximize the model accuracies~\cite{bishop2006pattern}.

The success of DL approaches in various areas, such as natural language processing~(NLP), computer vision, robotics, and database~\cite{lecun2015deep,wang2016database} in recent years have drawn the attention of the EL research community to a promising direction. Compared with traditional machine learning methods, DL is known to be capable of extracting task-specific features from raw data automatically through the learning process. In addition, the development of distributed representations enables DL models to process textual data directly~\cite{mikolov2013distributed,pennington2014glove,bojanowski2017enriching,joulin2017bag}. These properties of DL are highly desirable for EL frameworks.

\begin{figure}
	\centering
	\includegraphics[width=0.4\textwidth]{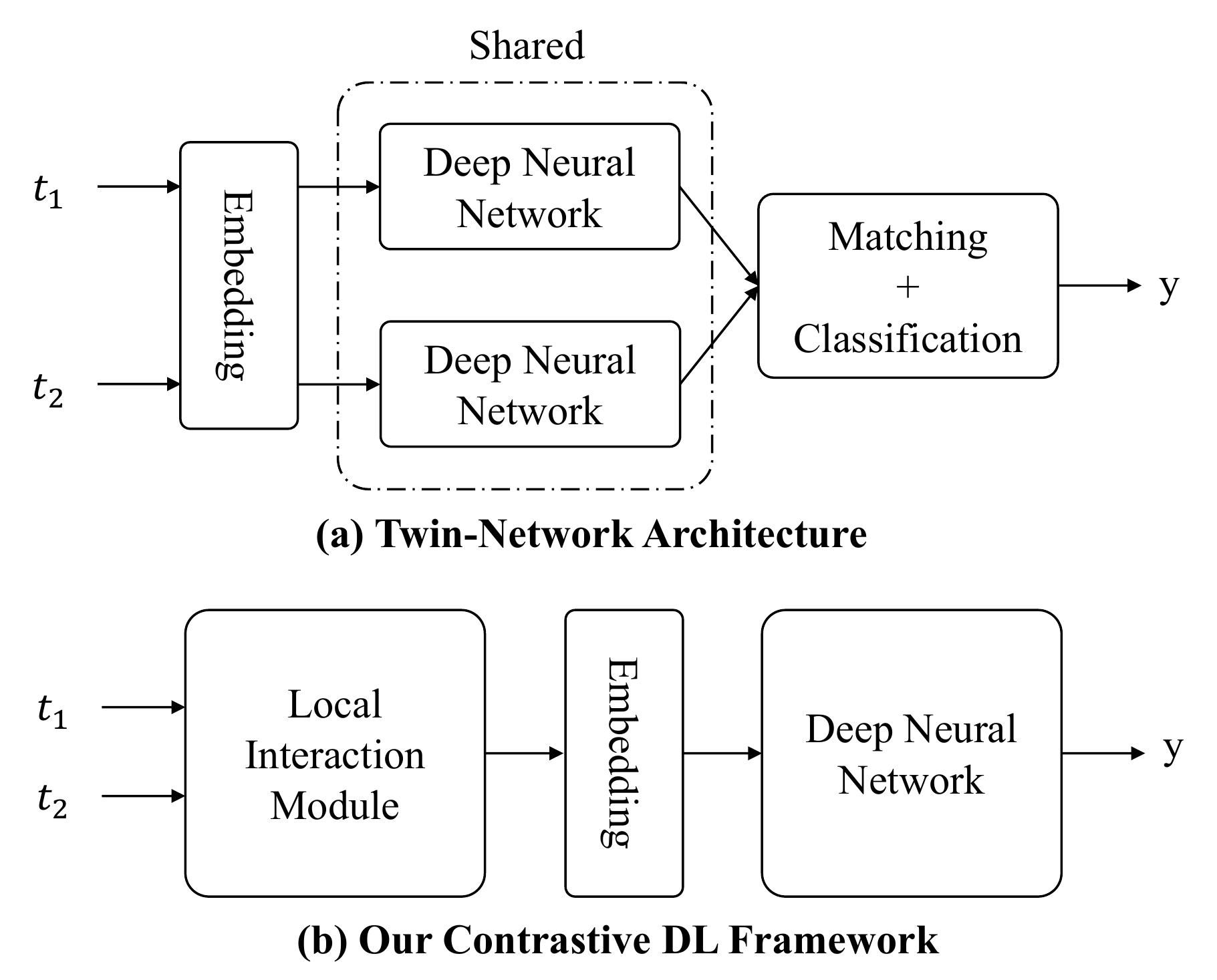}
	\caption{Two types of DL architectures for matching tasks. (a) The twin architecture employed by existing DL models for EL. The problems of applying the twin-network architecture on EL are analyzed in Section~\ref{sec:intro}. (b) Our proposed contrastive DL framework for EL followed by \textsc{CorDEL}. Details are described in Sections~\ref{sec:framework} and \ref{sec:local_interact}. The advantages of our framework are discussed in Section~\ref{sec:pros}.}
	\label{fig:architecture}
\end{figure}

Our work is not the first DL approach for EL. Existing DL methods for EL~\cite{ebraheem2018distributed,mudgal2018deep} employ the twin-network architecture in Figure~\ref{fig:architecture}(a), which is commonly used for other matching tasks in NLP in the literature. In NLP, the twin-network architecture is usually employed for semantic matching tasks such as question answering that require matching abstract text representations. However, semantic matching is not effective on many EL tasks. For example, in product EL tasks, the record pair (Black ink tank, Canon) and (Cyan ink tank, Canon), where the attributes are (Product title, Brand), is a non-match since they have different colors. However, the words representing different colors are semantically close to each other, making it difficult to distinguish this pair based on semantic matching. Another example is the record pair (Coca-Cola 12 fl oz 8 pack, Coca-Cola) and (Coca-Cola 12 fl oz 6 pack, Coca-Cola), where the only difference lies in the number of bottles in a pack. It is a non-match as well, even though words representing numbers have similar semantic meanings. In addition to these non-match cases, semantic matching could also fail on matches. For instance, the beer product record pair (Amber ale, Third Base Sports Bar \& Brewery) and (American red ale, Third Base Sports Bar \& Brewery) is a match. But the word `American' in one record is not semantically similar to any word in the other record, which may confuse semantic matching models. Besides these examples, recent studies have also shown that deep neural networks work like low-pass filters and have the effect of smoothing out small  differences~\cite{hamilton2018deep,nt2019revisiting}. Since the comparisons in the twin-network architecture is made after the records are projected onto the embedding space, small but crucial differences may be ignored, resulting in failures on EL tasks.

Because of these limitations of the twin-network architecture, existing DL models for EL do NOT show consistently improved performance over current non-DL machine learning models on various EL tasks. The fact that DL models may cause decreased performance in some cases hinders the use of these models for EL in practice.

In order to develop more effective and practical DL models for EL, we propose to jump out of the existing DL framework based on the twin-network architecture. Instead, we propose a new contrastive DL framework for EL, as shown in Figure~\ref{fig:architecture}(b)\footnote{Our contrastive DL framework does not correspond to the contrastive learning in the fields of deep metric learning and self-supervised learning. The ``contrastive'' here refers to contrasting one input to the other in the raw string level, as explained in Sections~\ref{sec:framework} and~\ref{sec:local_interact}.}. In contrast to the twin-network architecture, our framework is able to capture both syntactic and semantic signals. More importantly, our framework avoids the smoothing effect of deep neural networks and pays attention to subtle but critical differences. As an instantiation of this contrastive DL framework, we build a powerful DL model called \textsc{CorDEL} (C\textsc{o}nt\textsc{r}astive Deep Entity Linkage). Our contributions can be summarized in three aspects:
\begin{itemize}
    \item We propose a novel and generic contrastive DL framework for EL, as shown in Figure~\ref{fig:architecture}(b). Our contrastive framework addresses the limitations of the twin-network architecture in Figure~\ref{fig:architecture}(a) by capturing both syntactic and semantic signals and paying attention to subtle but critical differences between entities.
    \item We propose a powerful DL model called \textsc{CorDEL} (C\textsc{o}nt\textsc{r}astive Deep Entity Linkage) as an instantiation of our proposed contrastive DL framework, as illustrated in Figure~\ref{fig:CorDEL}. Concretely, we develop three variants of \textsc{CorDEL}, named \textsc{CorDEL}-Sum, \textsc{CorDEL}-Attention, and \textsc{CorDEL}-Context\_Attention.
    \item We perform extensive experiments on both public benchmark datasets and a large real-world dataset. \textsc{CorDEL} is able to outperform previous state-of-the-art models by 5.2$\%$ on public benchmark datasets. \textsc{CorDEL} also yields a 2.4$\%$ improvement over the current best DL model on the real-world dataset, while reducing 97.6$\%$ training parameters. In addition, \textsc{CorDEL} shows great stability over different runs. These results indicate that \textsc{CorDEL} is a reliable, efficient, and effective DL approach for EL.
\end{itemize}

\section{Related Work}

In this section, we discuss the twin-network architecture and review previous DL models for EL.

The twin-network architecture in Figure~\ref{fig:architecture}(a) has been widely applied on matching tasks in natural language processing~(NLP), such as paraphrase identification, question answering, automatic dialogue, and textual entailment~\cite{huang2013learning,gao2014modeling,shen2014learning,hu2014convolutional}. A notable property shared by these matching tasks is that they focus on semantic matching, \textit{i.e.}, the matching prediction is mainly determined by the semantic relations between two textual inputs. Employing twin-networks with pre-trained distributed representations~\cite{mikolov2013distributed,pennington2014glove,bojanowski2017enriching,joulin2017bag} suits these tasks well. It is because distributed representations are able to model semantic meanings. For example, words with similar semantic meanings have distributed representations with small distances. On the other hand, there are other matching tasks in NLP that do not fall into this category, such as relevance matching~\cite{guo2016deep}, paper citation matching~\cite{pang2016text}, etc. In these tasks, signals from syntactic matching are as important as those from semantic matching. In order to address these tasks, several DL matching models have been developed~\cite{lu2013deep,guo2016deep,pang2016text}. Typically, these models perform local interaction among two inputs and construct a matching histogram or comparison matrix. Deep neural networks are then applied on the matching histogram or comparison matrix to make predictions.

In the literature, most existing DL models for EL follow the twin-network architecture~\cite{mudgal2018deep,ebraheem2018distributed,nie2019deep,zhao2019auto}. \textsc{DeepMatcher}~\cite{mudgal2018deep} proposed a general twin-network template of DL models for EL, with four different instantiations: SIF, RNN, Attention, and Hybrid. \textsc{DeepER}~\cite{ebraheem2018distributed} shared high similarities with the SIF and RNN versions of \textsc{DeepMatcher} in terms of both the network architectures and performance. Seq2SeqMatcher~\cite{nie2019deep} augmented the twin-network architecture by proposed a sequence-to-sequence alignment layer, which shared certain similarities with the \textsc{DeepMatcher}-Attention. AutoEM~\cite{zhao2019auto} explored the transfer learning settings while still employing twin-network based DL models.

We have pointed out that the twin-network architecture is not suitable to EL tasks in Section~\ref{sec:intro}. In addition, because of the differences between EL and other tasks, existing non-twin DL matching models in NLP~\cite{lu2013deep,guo2016deep,pang2016text} could not be directly applied on EL tasks. In this work, we propose a novel and generic contrastive DL framework for EL, as shown in Figure~\ref{fig:architecture}(b). We propose a simple yet effective instantiation of this framework, named \textsc{CorDEL}.

\section{Method}\label{sec:method}

In this section, we first formally define the problem of EL in Section~\ref{sec:problem_def}. Then we propose our contrastive DL framework for EL in Section~\ref{sec:framework}. As an instantiation of the framework, we introduce \textsc{CorDEL}, a novel DL model for EL in Section~\ref{sec:local_interact}. We provide different powerful variants of \textsc{CorDEL} in Section~\ref{sec:CorDEL_variants}. Finally, we analyze the advantages of \textsc{CorDEL} in Section~\ref{sec:pros}.

\subsection{Problem Definition}\label{sec:problem_def}

We focus on EL that refers to the matching task between two data records. In detail, data records are saved by following a certain schema. That is, given an ordered set of pre-defined attributes, data are stored by putting its values under corresponding attributes. For example, the product record (Black ink tank, Canon) is saved with pre-defined attributes (Product title, Brand).

Formally, given pre-defined attributes $A_1, A_2, \ldots, A_m$, a data record $t$ can be represented as a tuple $(t[A_1], t[A_2], \ldots, t[A_m])$, where $t[A_i]$, $i=1,2,\ldots,m$ refers to the value of the attribute $A_i$ in the record $t$. In an EL dataset, all the records should have the same schema, that is, the same set of attributes in the same order. The EL task is to determine whether a pair of records $t_1$ and $t_2$, where $t_1 \neq t_2$, refer to the same real-world entity. In particular, it is formulated as a binary classification problem:
\begin{equation}
    y = F(t_1, t_2) \in \{0,1\},
\end{equation}
where $F$ represents a model for EL that outputs a binary prediction $y$. In practice, it is common to let $F$ first output a continuous number $y \in [0,1]$, and set a threshold to translate it into the binary classification result. The continuous output is called the matching score and can be interpreted as the likelihood of $t_1$ and $t_2$ being a match.

\subsection{Contrastive DL Framework}\label{sec:framework}

We first propose a novel and generic contrastive DL framework specially designed for EL, upon which we develop \textsc{CorDEL}. The framework is illustrated in Figure~\ref{fig:architecture}(b). We describe it component by component in this section.

\textbf{Local interaction module (LIM)}: In order to allow syntactic signal to be captured, our contrastive DL framework avoids projecting inputs into the embedding space at the beginning. Instead, it first employs a LIM to enable the two input records to interact with each other in the raw string level. The LIM compares and contrasts the input records in terms of string tokens, where the tokens can be characters, words, and phrases. After the LIM, all the string tokens from two input records are re-grouped, where each group captures specific syntactic signals. As a result, the outputs of the LIM are simply several groups of string tokens. Our instantiation, \textsc{CorDEL}, explores a simple LIM that simply separates the different words from the shared words appearing in both records, as introduced in Section~\ref{sec:local_interact}.

\textbf{Embedding}: With syntactic signals captured by the LIM through grouping, distributed embeddings~\cite{mikolov2013distributed,pennington2014glove,bojanowski2017enriching,joulin2017bag} of string tokens allow semantic signals to be taken into consideration by the following deep neural network. Therefore, our framework has an embedding layer after the LIM, which transforms each string token into a numeric vector embedding through distributed representations. The outputs of the embedding layer are thus sequences of vector embeddings corresponding to groups of string tokens. Note that, as the syntactic signals are encoded by the grouping, they will not be lost through the embedding layer. In other words, both syntactic and semantic signals are captured in the outputs of the embedding layer.

\textbf{Deep neural network}: Finally, a deep neural network is applied on top of the embedding layer to process both syntactic and semantic signals and make the prediction. As the inputs are sequences of vector embeddings, the deep neural network can be decomposed into three parts: sequence processing, information aggregation, and classification. First, for each group of vector embeddings, a sequence processing module is employed to summarize the information into a fixed-size vector representation. Next, the information from different groups needs to be aggregated, and then serves as inputs to a classification module.

The proposed contrastive DL framework is the first DL framework for EL that considers both syntactic and semantic signals. In the next section, we propose a powerful DL model as an instantiation of this framework, called \textsc{CorDEL}.

\subsection{An Instantiation-\textsc{CorDEL}}\label{sec:local_interact}

\begin{figure}
	\centering
	\includegraphics[width=0.45\textwidth]{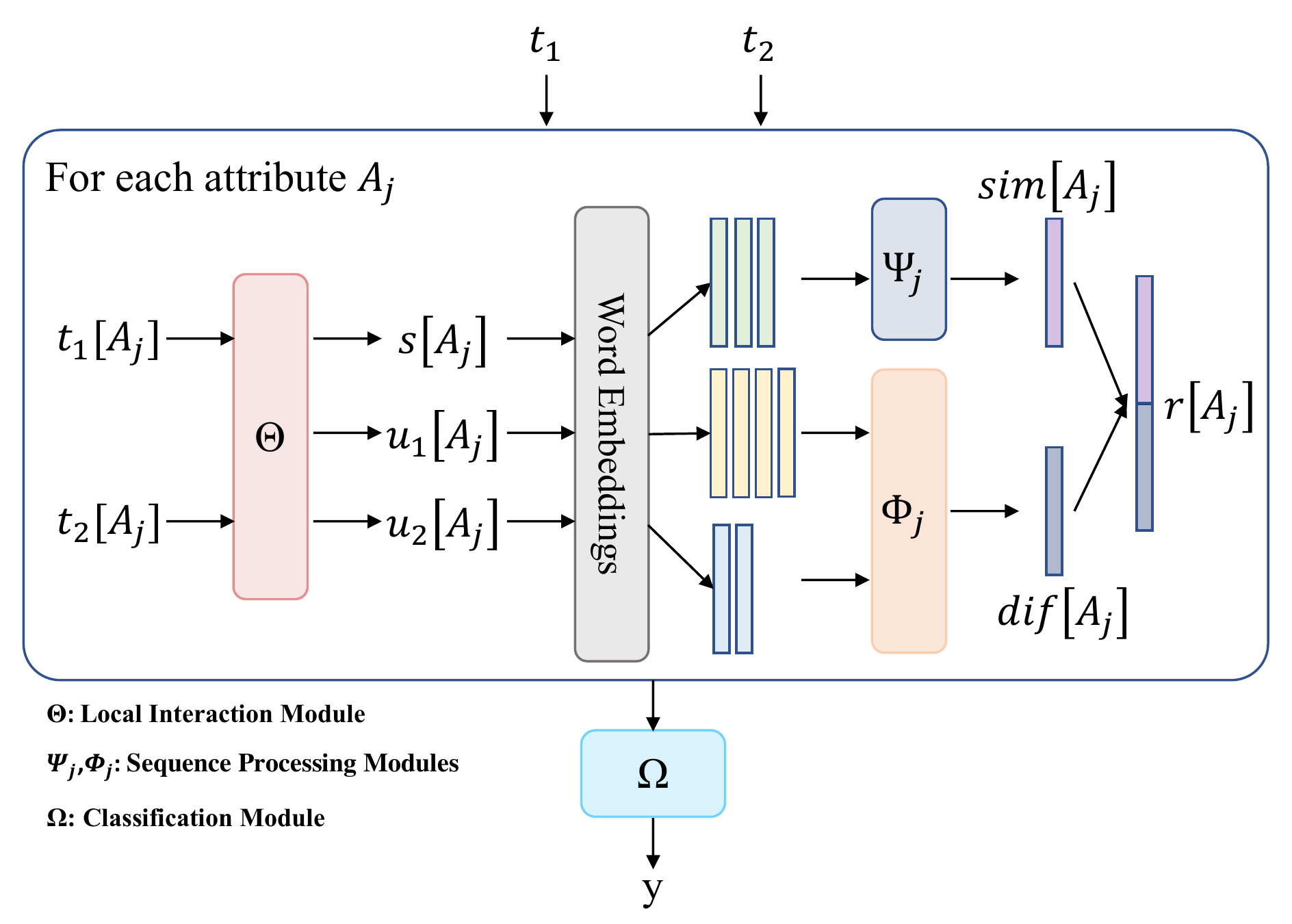}
	\caption{An illustration of our \textsc{CorDEL} described in Section~\ref{sec:local_interact}. It follows the proposed contrastive DL framework introduced in Section~\ref{sec:framework}. We provide different options for $\Psi_j$, $\Phi_j$, and $\Omega$ in Section~\ref{sec:CorDEL_variants}, leading to three variants of \textsc{CorDEL}.}
	\label{fig:CorDEL}
\end{figure}

An illustration of the proposed \textsc{CorDEL} (C\textsc{o}nt\textsc{r}astive Deep Entity Linkage) is provided in Figure~\ref{fig:CorDEL}. Specifically, under our proposed contrastive DL framework for EL, we develop a simple yet effective LIM followed by a carefully designed deep neural network.

\textbf{Local interaction module (LIM)}: The LIM of \textsc{CorDEL} is designed based on human intuition: given an input record pair, we tend to treat the differences between two records as signals for a non-match, and regard the common part as signals for a match. Therefore, our LIM simply separates the different words from the shared words appearing in both records. This results in re-clustering the tokens into three groups: two groups of unique words in either record, and one group of shared words. Specifically, the proposed LIM is achieved through simple set operations, as described below.

Formally, let $t_1$ and $t_2$ denote the input record pair, where $t_i=(t_i[A_1], t_i[A_2], \ldots, t_i[A_m])$, $i=1,2$, and each attribute value $t_i[A_j]$, $i=1,2$, $j=1,2,\ldots,m$, is a sequence of words. Our LIM $\Theta$ of \textsc{CorDEL} contrasts attribute-wise local tokens. For each attribute $A_j$, $j=1,2,\ldots,m$, the two sequences of words $t_1[A_j]$ and $t_2[A_j]$ are compared in terms of the exact matching between token sets. After  $\Theta$, all the tokens in $t_1[A_j]$ and $t_2[A_j]$ are distributed into three groups:
\begin{equation}
    (s[A_j], u_1[A_j], u_2[A_j]) = \Theta(t_1[A_j], t_2[A_j]),
\end{equation}
where $s[A_j]$ contains shared words appearing in both $t_1[A_j]$ and $t_2[A_j]$, and $u_i[A_j]$, $i=1,2$, includes the unique words that are only in $t_i[A_j]$. In other words, the comparison step $\Theta$ can be written as
\begin{align}
    &s[A_j] = t_1[A_j] \cap t_2[A_j], \nonumber \\
    &u_1[A_j] = t_1[A_j] \setminus s[A_j], \nonumber \\
    &u_2[A_j] = t_2[A_j] \setminus s[A_j].
\end{align}

\textbf{Embedding}: Accordingly, \textsc{CorDEL} employs pre-trained word embeddings to transform the outputs of $\Theta$ into word embeddings. Without loss of clarity, the same notations $(s[A_j], u_1[A_j], u_2[A_j])$ are used to denote the corresponding three sequences of word embeddings.

\textbf{Deep neural network}: We introduce the corresponding deep neural network in the order of sequence processing, information aggregation, and classification.

(1) \textit{Sequence processing}: For each attribute $A_j$, two sequence processing modules, $\Psi_j$ and $\Phi_j$, are used to generate an attribute similarity representation vector $sim[A_j]$ and an attribute difference representation vector $dif[A_j]$ from $(s[A_j], u_1[A_j], u_2[A_j])$, respectively:
\begin{align}
    &sim[A_j] = \Psi_j(s[A_j]), \label{eqn:summarize1} \\
    &dif[A_j] = \Phi_j(u_1[A_j], u_2[A_j]). \label{eqn:summarize2}
\end{align}
Note that we use one sequence processing module $\Phi_j$ to process two groups $u_1[A_j]$ and $u_2[A_j]$ instead of two distinct ones. This is because both groups include different words, which can be viewed as one group as well. Here, the attribute similarity representation vector $sim[A_j]$ encodes information from shared words under the attribute $A_j$ in both records, serving as evidence that supports the prediction of the input record pair as a match. On the contrary, the attribute difference representation vector $dif[A_j]$ encodes information from different words under the attribute $A_j$ in either record, supporting the opposite prediction.

(2) \textit{Information aggregation}: In order to aggregate information, \textsc{CorDEL} concatenates $sim[A_j]$ and $dif[A_j]$ as the attribute representation vector $r[A_j]$:
\begin{equation}
     r[A_j] = \text{Concat}(sim[A_j], dif[A_j]).
\end{equation}

(3) \textit{Classification}: Finally, a classification module $\Omega$ takes all $m$ attribute representation vectors as inputs and performs a binary classification task:
\begin{equation}\label{eqn:classify}
    y = \Omega(r[A_1],r[A_2],\ldots,r[A_m]) \in [0,1],
\end{equation}
where $y$ is the predicted matching score. A threshold can be set to translate the matching scores into binary classification results. The classification module $\Omega$ has to merge $m$ vectors first and makes the prediction. In DL, it is common to let $\Omega$ output two numbers, use the Softmax function to normalize them, and treat one of them as the $y$ in Eqn.~(\ref{eqn:classify})~\cite{bishop2006pattern}. With the true label $y^*$ from the training dataset, \textsc{CorDEL} can be trained with the cross-entropy loss through back-propagation~\cite{lecun1998gradient}.

\subsection{Variants of \textsc{CorDEL}}\label{sec:CorDEL_variants}

In this section, we provide variants of \textsc{CorDEL} by specifying $\Psi_j$ in Eqn.~(\ref{eqn:summarize1}), $\Phi_j$ in Eqn.~(\ref{eqn:summarize2}), and $\Omega$ in Eqn.~(\ref{eqn:classify}). In particular, $\Psi_j$ and $\Phi_j$ are each required to take in one and two variable-length sequence of word embeddings and produce a fixed-size vector, respectively. And $\Omega$ has $m$ fixed-size vectors as inputs and performs a two-way classification.

\textbf{\textsc{CorDEL}-Sum:} In order to demonstrate the effectiveness of our proposed \textsc{CorDEL}, we build \textsc{CorDEL}-Sum, an extremely simple variant of \textsc{CorDEL}.

\textsc{CorDEL}-Sum employs summation followed by a one-layer mutlilayer perceptron (MLP) for both $\Psi_j$ and $\Phi_j$. Summation, although without any training parameters, is a powerful process in DL models for classification tasks~\cite{joulin2017bag,xu2019powerful}. The one-layer MLP is used to perform dimension reduction, which avoids having an excessive number of parameters in the following classification module $\Omega$. Specifically, we have
\begin{align}
    &sim[A_j] = \Psi_j(s[A_j]) = \sigma(W^{\Psi_j}\cdot \sum_{s \in s[A_j]} s), \nonumber \\
    &dif[A_j] = \Phi_j(u_1[A_j], u_2[A_j]) = \sigma(W^{\Phi_j}\cdot \sum_{u \in u_1[A_j] \cup u_2[A_j]} u), \nonumber
\end{align}
where $W^{\Psi_j}$ and $W^{\Phi_j}$ represent corresponding one-layer MLPs, and $\sigma$ refers to an activation function. The bias terms are omitted. In particular, $\Phi_j$ sums all the input word embeddings from both sequences of difference words. It is worth noting that the one-layer MLPs are independent for each attribute $A_j$, leading to $2m$ one-layer MLPs in total. 

Afterwards, $\Omega$ of \textsc{CorDEL}-Sum is simply implemented as a concatenation of $m$ input vectors followed by a two-layer MLP with two output units:
\begin{equation}\label{eqn:sum-classify}
    y = \text{MLP}(\text{Concat}(r[A_1],r[A_2],\ldots,r[A_m])).
\end{equation}

\textsc{CorDEL}-Sum is extremely light-weight yet powerful. The training parameters only lie in $2m$ one-layer MLPs plus a two-layer MLP. As shown in Section~\ref{sec:experiments}, \textsc{CorDEL}-Sum achieves significantly improved performance over current non-DL and DL models. The success of \textsc{CorDEL}-Sum demonstrates the power of our proposed \textsc{CorDEL}.

\textbf{Attention-based \textsc{CorDEL}:} Despite the effectiveness of \textsc{CorDEL}-Sum, using summation to perform sequence processing may limit the performance in some cases, as summation gives equal importance to each word in the sequence. This contradicts with the intuition that words in $s[A_j]$ and $(u_1[A_j], u_2[A_j])$ should contribute differently to $sim[A_j]$ and $dif[A_j]$, respectively. Therefore, we explore attention-based modules for $\Psi_j$, $\Phi_j$, and $\Omega$ to further enhance our \textsc{CorDEL}. The attention mechanism is able to perform a weighted summation over word embeddings, giving larger weights to more important words. 

The attention mechanism has been widely used in DL models for various computer vision and NLP tasks~\cite{yang2016hierarchical,vaswani2017attention,wang2018non,devlin2019bert,wang2020non,wang2020icapsnets,liu2020global}. In general, the attention mechanism has three parts of inputs: a query vector $q \in \mathbb{R}^{d_1}$, $n$ key vectors that form a matrix $K=[k_1,k_2,\ldots,k_n] \in \mathbb{R}^{d_1 \times n}$, and $n$ value vectors that form a matrix $V=[v_1,v_2,\ldots,v_n] \in \mathbb{R}^{d_2 \times n}$. Notably, the dimension of the query vector and key vectors are the same, and key vectors and value vectors have a one-to-one correspondence. The attention mechanism~\cite{vaswani2017attention} is defined as
\begin{equation}
    o = V \cdot \text{Softmax}(\frac{K^T \cdot q}{\sqrt{d_1}}) \in \mathbb{R}^{d_2}.
\end{equation}

In order to use the attention mechanism, we need to specify where the $q$, $K$, and $V$ come. With different choices, we develop two attention-based variants of \textsc{CorDEL}, named \textsc{CorDEL}-Attention and \textsc{CorDEL}-Context\_Attention. They differ in $\Phi_j$, while having the same $\Psi_j$ and $\Omega$.

We describe the shared $\Psi_j$ first. To simplify the notations, let $s[A_j]=[x_1,x_2,\ldots,x_n] \in \mathbb{R}^{d \times n}$ denote the inputs to $\Psi_j$. Note that $n$ can be any number so that $s[A_j]$ is a variable-length sequence of embeddings. The $K$ and $V$ in the attention mechanism are computed from the inputs $s[A_j]$ through $K = W^k X$ and $V = W^v X$, where $W^k \in \mathbb{R}^{d_1 \times d}$ and $W^v \in \mathbb{R}^{d_2 \times d}$ are training parameters. Meanwhile, the query vector $q$ is simply randomly initialized and tuned during training~\cite{yang2016hierarchical}.

In terms of $\Phi_j$, both \textsc{CorDEL}-Attention and \textsc{CorDEL}-Context\_Attention follow a sub-twin architecture, that is, two attention mechanisms with shared training parameters are applied on $u_1[A_j]$ and $u_2[A_j]$, respectively. And the output of $\Phi_j$ is the summation of the outputs from the two attention mechanisms. Like the attention mechanism in in $\Psi_j$, the attention mechanisms on $u_i[A_j]$ compute $K$ and $V$ from the inputs $u_i[A_j]$. However, \textsc{CorDEL}-Attention and \textsc{CorDEL}-Context\_Attention have different choises on $q$. \textsc{CorDEL}-Attention employ the attention with trainable $q$ as in $\Phi_j$, while \textsc{CorDEL}-Context\_Attention uses the output of $\Phi_j$ as $q$, \textit{i.e.}, $q=sim[A_j]$.

The motivation of \textsc{CorDEL}-Attention is straightforward. As the attention mechanism may be more powerful than summation in some cases, \textsc{CorDEL}-Attention uses attention mechanisms with trainable $q$ to replace summations in \textsc{CorDEL}-Sum. On the other hand,  \textsc{CorDEL}-Context\_Attention uses $sim[A_j]$ to guide the attention mechanisms that generate $dif[A_j]$. The motivation is that $sim[A_j]$ may contain contextual information, and can be useful in determining the importance of words in $u_1[A_j]$ and $u_2[A_j]$. For example, the model can figure out that the domain of the input records is music. Within this context, words indicating the versions of the music records, such as `live' and `remix', should be paid more attention to.

Both \textsc{CorDEL}-Attention and \textsc{CorDEL}-Context\_Attention exploit self-attention~\cite{vaswani2017attention,devlin2019bert} in $\Omega$. By having $m$ query vectors, the attention mechanism is able to transform a sequence of embeddings into another sequence of embeddings with the same length~\cite{vaswani2017attention,devlin2019bert}. In particular, let $R= [r[A_1],r[A_2],\ldots,r[A_m]]$, we have $Q= W^Q R$, $K = W^k R$, and $V = W^v R$, where $W^q$, $W^k$, and $W^v$ are training parameters. Using self-attention to replace the concatenation in Eqn.~(\ref{eqn:sum-classify}) allows explicit cross-attribute interaction, leading to improved performance in some cases, as shown in Section~\ref{sec:experiments}.

\subsection{Analysis of \textsc{CorDEL}}\label{sec:pros}

We analyze the \textsc{CorDEL} and demonstrate its advantages. In particular, we demonstrate how it appropriately addresses the problems of existing DL models for EL.

By taking the LIM $\Theta$, \textsc{CorDEL} takes syntactic signals from raw strings into consideration. Meanwhile, semantic signals are still captured through word embeddings. On one hand, $\Theta$ helps \textsc{CorDEL} avoid mistakes caused by the fact that some semantically similar words are the key evidence for the prediction of a non-match. Taking the example of (Coca-Cola 12 fl oz 8 pack, Coca-Cola) and (Coca-Cola 12 fl oz 6 pack, Coca-Cola), the words `8' and `6' will be put into the groups of unique words in either record, and encoded by the attribute difference representation vector $dif[A_j]$. In the case that `8' and `6' have similar word embeddings as they are semantically close, \textsc{CorDEL} is still able to know that there is a numeric difference between the two input records, while the twin networks are not sensitive to such a difference. On the other hand, \textsc{CorDEL} is also effective in the case that semantically different but unimportant words make the model fail to identify a true match. As the final classifier takes both the attribute similarity representation vector $sim[A_j]$ and the attribute difference representation vector $dif[A_j]$ into consideration, \textsc{CorDEL} is able to determine whether the captured differences serve as important evidence for the prediction.

In addition, \textsc{CorDEL} is unaffected by the smoothing effect of deep neural networks. The differences are isolated from the common parts of the input record pair and processed separately. Therefore, no matter how small the differences are, \textsc{CorDEL} is capable of capturing them.

To summarize, unlike existing DL models for EL, \textsc{CorDEL} is able to identify subtle but critical differences between input records, which is a fundamental requirement for solving EL.

\section{Experimental Studies}\label{sec:experiments}

In this section, we conduct thorough experiments to evaluate our proposed \textsc{CorDEL} and show its superiority in the following aspects:
\begin{itemize}
    \item On public benchmark datasets, \textsc{CorDEL} outperforms existing non-DL and DL models on all types of EL tasks. In particular, \textsc{CorDEL} is the first DL approach with consistent and significant improvements over the non-DL approach on all three types of EL tasks.
    
    \item On a real-world dataset, \textsc{CorDEL} achieves better performance over existing DL models in terms of two practical evaluation metrics. In addition, \textsc{CorDEL} demonstrates significantly improved stability over independent training runs, which is highly desired in practice.
    
    \item \textsc{CorDEL} is a much more efficient DL approach in terms of required computational resources.
\end{itemize}

\subsection{Experimental Setup}

We describe the models and configurations used in our experiments.

\textbf{Baselines:} We select non-DL and DL baselines for comparison.
\begin{itemize}
    \item The non-DL baseline is Magellan~\cite{konda2016magellan}, the state-of-the-art machine learning based approach for EL. In particular, Magellan selects the best classifier from decision tree, random forest, Naive Bayes, support vector machine and logistic regression. The features used in Magellan are designed by experts.
    
    \item The DL baseline is \textsc{DeepMatcher}~\cite{mudgal2018deep}, which represents a wide range of twin-network based DL models for EL. \textsc{DeepMatcher} has four versions, named SIF, RNN, Attention, and Hybrid, with increasing complexity. \textsc{DeepER}~\cite{ebraheem2018distributed} and Seq2SeqMatcher~\cite{nie2019deep} can be regarded as extensions of \textsc{DeepMatcher}. \textsc{DeepMatcher} has been made publicly available as a Python package.
\end{itemize}

\textbf{\textsc{CorDEL}:} We evaluate \textsc{CorDEL}-Sum, \textsc{CorDEL}-Attention, and \textsc{CorDEL}-Context\_Attention in our experiments. The details are provided below.

\textit{Word Embeddings}: For fair comparison, the distributed representations used to transform words into word embeddings are 300-dimensional pretrained FastText embeddings~\cite{joulin2017bag}, which is the same as \textsc{DeepMatcher}~\cite{mudgal2018deep}. The embeddings are not fine-tuned during training.

\textit{Training}: \textsc{CorDEL} is trained through the Adam optimizer~\cite{kingma2014adam} with a learning rate of 0.0001. The training batch size is set to 64 for public datasets and 256 for the real-world dataset.

\textit{\textsc{CorDEL}-Sum}: As described in Section~\ref{sec:CorDEL_variants}, the training parameters of \textsc{CorDEL}-Sum only lie in $2m$ one-layer MLPs plus a two-layer MLP, where $m$ is the number of attributes in the dataset. The output dimension is set to 64 for the $2m$ one-layer MLPs. The dimension of the hidden layer in the two-layer MLP is set to 256.

\textit{\textsc{CorDEL}-Attention \& \textsc{CorDEL}-Context\_Attention}: As introduced in Section~\ref{sec:CorDEL_variants}, we only need to specify the dimension of training parameters in the attention mechanism, \textit{i.e.}, $d$, $d_1$ and $d_2$. In particular, $d$ depends on the dimension of word embeddings so that is 300 as indicated above. In the attention mechanism with a trainable query vector $q$, we $d_1$ to 4, a small number to prevent over-fitting. In the context-attention and self-attention modules, $d_1$ is set to 64. In all cases, $d_2$ is set to 64.

\subsection{Datasets}~\label{sec:datasets}

\begin{table}
	\centering
	\caption{Statistics of public benchmark datasets provided by~\cite{mudgal2018deep} and our real-world music dataset. The rightmost three columns, \textit{i.e.} \#Pairs, \#Matches, \#Attrs, correspond to the number of record pairs, matches, attributes in the dataset, respectively.}
	\label{table:public_datasets}
	\resizebox{0.48\textwidth}{!}{
	\begin{tabular}{clcccc}
		\toprule
		Type & Dataset & Domain & \#Pairs & \#Matches & \#Attrs \\
		\midrule
	    \multirow{7}{*}{Structured}
	    & BeerAdvo-RateBeer & beer & 450 & 68 & 4 \\
		& iTunes-Amazon$_1$ & music & 539 & 132 & 8 \\
		& Fodors-Zagats & restaurant & 946 & 110 & 6 \\
		& DBLP-ACM$_1$ & citation & 12,363 & 2,220 & 4 \\
		& DBLP-Scholar$_1$ & citation & 28,707 & 5,347 & 4 \\
		& Amazon-Google & software & 11,460 & 1,167 & 3 \\
		& Walmart-Amazon$_1$ & electronics & 10,242 & 962 & 5 \\
		\midrule
		Textual & Abt-Buy & product & 9,575 & 1,028 & 3 \\
		\midrule
		\multirow{4}{*}{Dirty}
		& iTunes-Amazon$_2$ & music & 539 & 132 & 8 \\
		& DBLP-ACM$_2$ & citation & 12,363 & 2,220 & 4 \\
		& DBLP-Scholar$_2$ & citation & 28,707 & 5,347 & 4 \\
		& Walmart-Amazon$_2$ & electronics & 10,242 & 962 & 5 \\
		\midrule
		Real-World & Amazon-Wikipedia & music & $\sim$0.4M & $\sim$0.2M & 10 \\
		\bottomrule
	\end{tabular}}
\end{table}

\begin{table*}
	\centering
	\caption{Comparisons between \textsc{CorDEL} and baselines on structured EL datasets from ~\cite{mudgal2018deep} in terms of the $F_1$ score. ``C\_Attention'' is short for ``Context\_Attention''. The best performance is highlighted with boldface. If \textsc{CorDEL} achieves the best performance, we mark the best results obtained by baselines with underlines, and vice versa. In particular, when \textsc{CorDEL} sets the new state-of-the-art record, the relative improvement rate against the previous best performance is computed. 
	}
	\label{table:public_structured}
	\begin{tabularx}{\textwidth}{p{2.55cm}c*{4}{>{\centering\arraybackslash}X}*{3}{>{\centering\arraybackslash}X}}
		\toprule
	    & \multirow{2}{*}{Magellan~\cite{konda2016magellan}} & \multicolumn{4}{c}{\textsc{DeepMatcher}~\cite{mudgal2018deep}} & \multicolumn{3}{c}{\textsc{CorDEL} (Ours)} \\
	    \cmidrule(lr){3-6} \cmidrule(l){7-9}
		Dataset & & SIF & RNN & Attention & Hybrid & Sum & Attention & C\_Attention \\
		\midrule
		BeerAdvo-RateBeer & \underline{78.8} & 58.1 & 72.2 & 64.0 & 72.7 & \textbf{88.9} $\uparrow_{12.8\%}$ & 85.7 & 86.7 \\
		iTunes-Amazon$_1$ & \underline{91.2} & 81.4 & 88.5 & 80.8 & 88.0 & \textbf{100.0} $\uparrow_{9.6\%}$ & 96.3 & 94.5 \\
		Fodors-Zagats & \underline{100.0} & \underline{100.0} & \underline{100.0} & 82.1 & \underline{100.0} & \textbf{100.0} $\uparrow_{0.0\%}$ & \textbf{100.0} & \textbf{100.0} \\
		DBLP-ACM$_1$ & \underline{98.4} & 97.5 & 98.3 & \underline{98.4} & \underline{98.4} & \textbf{99.2} $\uparrow_{0.8\%}$ & 98.9 & 98.8 \\
		DBLP-Scholar$_1$ & 92.3 & 90.9 & 93.0 & 93.3 & \textbf{94.7} & \underline{94.0} & 93.4 & 93.5 \\
		Amazon-Google & 49.1 & 60.6 & 59.9 & 61.1 & \underline{69.3} & \textbf{70.2} $\uparrow_{1.3\%}$ & 68.8 & 68.1 \\
		Walmart-Amazon$_1$ & \underline{71.9} & 65.1 & 67.6 & 50.0 & 66.9 & 68.7 & \textbf{72.7} $\uparrow_{1.1\%}$ & 70.9 \\
		\midrule
		Average $F_1$ & 83.1 & 79.1 & 82.8 & 75.7 & \underline{84.3} & \textbf{88.7} $\uparrow_{5.2\%}$ & 88.0 & 87.5 \\
		\bottomrule
	\end{tabularx}
\end{table*}

\begin{table*}
	\centering
	\caption{Comparisons between \textsc{CorDEL} and baselines on textual EL datasets from ~\cite{mudgal2018deep} in terms of the $F_1$ score.}
	\label{table:public_textual}
	\begin{tabularx}{\textwidth}{p{2.55cm}c*{4}{>{\centering\arraybackslash}X}*{3}{>{\centering\arraybackslash}X}}
		\toprule
	    & \multirow{2}{*}{Magellan~\cite{konda2016magellan}} & \multicolumn{4}{c}{\textsc{DeepMatcher}~\cite{mudgal2018deep}} & \multicolumn{3}{c}{\textsc{CorDEL} (Ours)} \\
	    \cmidrule(lr){3-6} \cmidrule(l){7-9}
		Dataset & & SIF & RNN & Attention & Hybrid & Sum & Attention & C\_Attention \\
		\midrule
		Abt-Buy & 43.6 & 35.1 & 39.4 & 56.8 & \underline{62.8} & 58.2 & \textbf{64.9} $\uparrow_{3.3\%}$ & 61.3 \\
		\bottomrule
	\end{tabularx}
\end{table*}

\begin{table*}
	\centering
	\caption{Comparisons between \textsc{CorDEL} and baselines on dirty EL datasets from ~\cite{mudgal2018deep} in terms of the $F_1$ score.}
	\label{table:public_dirty}
	\begin{tabularx}{\textwidth}{p{2.55cm}c*{4}{>{\centering\arraybackslash}X}*{3}{>{\centering\arraybackslash}X}}
		\toprule
	    & \multirow{2}{*}{Magellan~\cite{konda2016magellan}} & \multicolumn{4}{c}{\textsc{DeepMatcher}~\cite{mudgal2018deep}} & \multicolumn{3}{c}{\textsc{CorDEL} (Ours)} \\
	    \cmidrule(lr){3-6} \cmidrule(l){7-9}
		Dataset & & SIF & RNN & Attention & Hybrid & Sum & Attention & C\_Attention \\
		\midrule
		iTunes-Amazon$_2$ & 46.8 & 66.7 & \underline{79.4} & 63.6 & 74.5 & 82.1 & 78.0 & \textbf{82.4} $\uparrow_{3.8\%}$\\
		DBLP-ACM$_2$ & 91.9 & 93.7 & 97.5 & 97.4 & \textbf{98.1} & \underline{97.0} & 96.3 & 96.8 \\
		DBLP-Scholar$_2$ & 82.5 & 87.0 & 93.0 & 92.7 & \textbf{93.8} & \underline{91.9} & 89.0 & 89.9 \\
		Walmart-Amazon$_2$ & 37.4 & 43.2 & 39.6 & \textbf{53.8} & 46.0 & 48.3 & 50.1 & \underline{51.2} \\
		\midrule
		Average $F_1$ & 64.7 & 77.4 & 76.9 & 75.7 & \underline{78.1} & 79.8 & 78.4 & \textbf{80.1} $\uparrow_{2.6\%}$ \\
		\bottomrule
	\end{tabularx}
\end{table*}

Experiments are performed on public benchmark datasets and a real-world dataset. Various evaluation metrics are used.

\textbf{Public Benchmark Datasets:} We conduct experiments on the public datasets provided by~\cite{mudgal2018deep}. These public datasets cover a wide range of EL tasks in different domains. In particular, they represent three types of EL tasks.
\begin{itemize}
    \item \textbf{Structured EL}: In a structured EL dataset, the records in a pair have relatively clean and aligned attribute values. In addition, the number of tokens in an attribute value is usually limited.
    
    \item \textbf{Textual EL}: As indicated by the name, a textual EL dataset has long textual data as attribute values.
    
    \item \textbf{Dirty EL}: A dirty EL dataset differs from a structured EL dataset in the aspect that the attribute values may be mistakenly disposed. The value of one attribute could appear as part of the value of another attribute.
\end{itemize}

In total, there are 7 structured, 1 textual, and 4 dirty EL datasets. The statistics of these datasets are provided in Table~\ref{table:public_datasets}. Following~\cite{mudgal2018deep}, we divide each dataset into training, validation, and evaluation splits with the ratio of 3:1:1.

In the experiments on these public datasets, we follow \cite{mudgal2018deep} to employ the $F_1$ score as the evaluation metric, which allows the direct comparison between our proposed \textsc{CorDEL} and baselines. Note that, according to Eqn.~(\ref{eqn:classify}), the output of \textsc{CorDEL} is a matching score $y \in [0,1]$. A threshold has to be set to transform the matching score into a binary classification result. As with \cite{mudgal2018deep}, we set the threshold to 0.5 to compute $F_1$.

It is easy and beneficial for research purpose to classify current public benchmark datasets for EL tasks~\cite{mudgal2018deep} according to such categorization. However, real-world EL datasets may be a mixture of the three types. Therefore, a general approach for EL that is able to achieve good performance consistently on any type of EL task is highly desired in practice.

\textbf{Real-world Dataset:} We collect a real-world EL dataset in the music domain. Specifically, music records are crawled and sampled from Amazon and Wikipedia \cite{zhucollective}. That is, in a record pair $t_1$ and $t_2$ from this dataset, $t_1$ is from Amazon and $t_2$ is from Wikipedia. We have 10 attributes describing basic information about the music track records.
In order to obtain the training dataset, we sample 0.4 million record pairs involving 822,276 distinct entities and employ a noisy strong key to label them. Meanwhile, the testing dataset contains record pairs that are manually labelled by human annotators, ensuring that the evaluation is accurate.

We adopt more comprehensive and practical evaluation metrics for experiments on this real-world dataset: Area Under the Precision-Recall Curve (PRAUC) and Recall when Precision=95$\%$ (R@P=95$\%$). The $F_1$ score evaluates the model when a chosen threshold. In contrast, PRAUC summarizes the model performance with all thresholds. In addition, as most EL datasets are imbalanced, PRAUC is known to be more suitable for evaluating binary classifiers on imbalanced datasets~\cite{saito2015precision}. R@P=95$\%$ is a practical evaluation metric for EL. Data integration typically has the requirement for high precision. That is because a low-precision approach for EL would result in wrongly merges records, causing unrecoverable data loss.

\subsection{Results on Public Datasets}

We compare \textsc{CorDEL} with baselines on three types of public EL datasets separately. The results of baselines are provided by~\cite{mudgal2018deep}.

\textbf{Structured EL}: Results on the 7 structured EL datasets are reported in Table~\ref{table:public_structured}. All versions of \textsc{CorDEL} improve the performance by a large margin in terms of the average $F_1$ score. Notably, \textsc{CorDEL}-Sum achieves the state-of-the-art performance on 5 out of 7 datasets. On DBLP-Scholar$_1$, \textsc{CorDEL}-Sum is the second best model while the best model \textsc{DeepMatcher}-Hybrid has 32x more parameters, as shown in Section~\ref{sec:efficiency}. On Walmart-Amazon$_1$, \textsc{CorDEL}-Sum outperforms all versions of the DL baseline. In addition, \textsc{CorDEL}-Attention achieves the best result on Walmart-Amazon$_1$, being the only DL model that beats the non-DL baseline.
    
\textsc{CorDEL}-Sum yields a 5.2$\%$ improvement over the previous state-of-the-art model in terms of the average $F_1$ scores. While existing DL models can only achieve competitive performance with non-DL models, \textsc{CorDEL} is the first DL approach that demonstrates the advantages of DL on structured EL tasks.
    
\textbf{Textual EL}: Table~\ref{table:public_textual} shows the results on the textual EL dataset Abt-Buy. It is a valid concern that the local string comparison step $\Theta$ breaks the long textual attribute values, such as sentences and paragraphs, which might harm the performance of \textsc{CorDEL} on textual EL tasks. However, experimental results indicate that our proposed \textsc{CorDEL} remains powerful. Moreover, \textsc{CorDEL}-Attention sets the new state-of-the-art record, increasing the best $F_1$ score by 3.3$\%$.
    
\textbf{Dirty EL}: Table~\ref{table:public_dirty} provides the results on the 4 dirty EL datasets. The advantage of using DL models for dirty EL tasks is inherited by \textsc{CorDEL}. While only obtaining the best results on 1 out of 4 datasets by \textsc{CorDEL}-Context\_Attention, \textsc{CorDEL} achieves the best average $F_1$ score. Particularly, \textsc{CorDEL}-Context\_Attention improves the best average $F_1$ score by 2.6$\%$. It indicates that \textsc{CorDEL} is more robust to different datasets.

To conclude, \textsc{CorDEL} is the first DL approach that yields consistently and significantly improved performance on various datasets for different types of EL tasks, serving as a general DL approach for EL.

\subsubsection{Case Studies}\label{sec:case_study}

\begin{figure}
	\centering
	\includegraphics[width=0.4\textwidth]{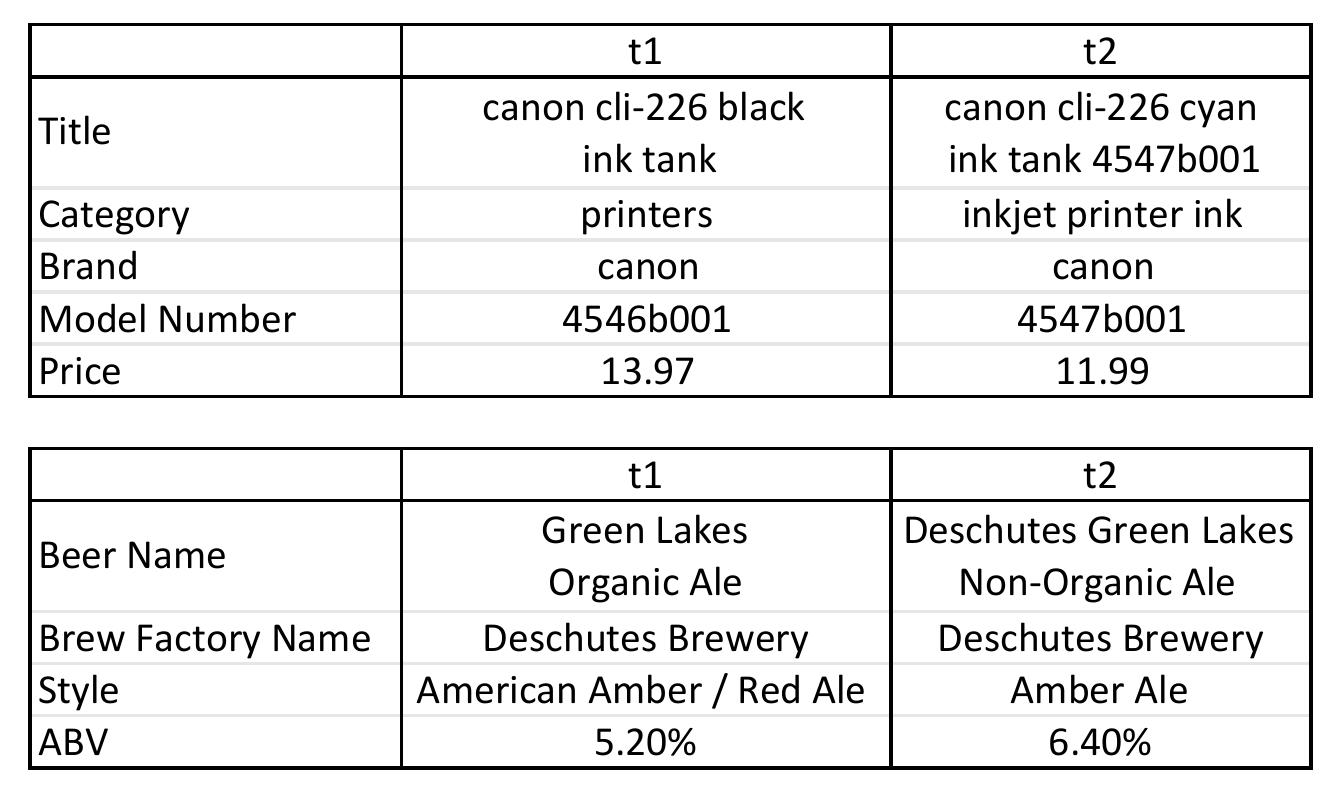}
	\caption{Case studies on public benchmark datasets. The top example is from Walmart-Amazon$_1$, the bottom example is from BeerAdvo-RateBeer. Both of them are non-matches, with subtle but critical differences. \textsc{CorDEL} makes the correct prediction in both cases, while \textsc{DeepMatcher} fails.}
	\label{fig:cases}
\end{figure}

We perform case studies to show why \textsc{CorDEL} achieves better performance. Specifically, we examine examples in the testing dataset, where \textsc{CorDEL} makes the correct prediction but \textsc{DeepMatcher} fails. Figure~\ref{fig:cases} provides two representative examples from Walmart-Amazon$_1$ and BeerAdvo-RateBeer, respectively. Both of them are non-matches, with subtle but critical differences. However, \textsc{DeepMatcher} identifies them as matches, indicating its inability to capture those subtle but critical differences between input records. On the contrary, as discussed in Section~\ref{sec:pros}, \textsc{CorDEL} has an outstanding ability to handle such cases.

\begin{table*}
	\centering
	\caption{Comparisons between \textsc{CorDEL} and baselines on a real-world dataset in terms of Area Under the Precision-Recall Curve (PRAUC), Recall when Precision=95$\%$ (R@P=95$\%$), and the number of training parameters in total (\#Params). The relative improvement rates against the previous best model, \textsc{DeepMatcher}-Hybrid, are computed.}
	\label{table:real_world}
	\begin{tabular}{lccc}
		\toprule
		Model & PRAUC & R@P=95$\%$ & \#Params \\
		\midrule
		\textsc{DeepMatcher}-SIF & 88.1 $\pm$ 2.9 & 43.5 $\pm$ 17.0 & 728,802 \\
		\textsc{DeepMatcher}-Hybrid & 90.5 $\pm$ 1.9 & 52.7 $\pm$ 25.1 & 22,151,812 \\
		\midrule
		\textsc{CorDEL}-Sum & 91.6 $\pm$ 0.3 & \textbf{68.2} $\pm$ \textbf{2.4} $\uparrow_{29.4\%}$ & 713,730 \\
		\textsc{CorDEL}-Attention & \textbf{92.7} $\pm$ \textbf{0.3} $\uparrow_{2.4\%}$ & 67.8 $\pm$ 1.3 & \textbf{522,850} $\downarrow {97.6\%}$\\
		\bottomrule
	\end{tabular}
\end{table*}


\subsection{Results on the Real-world Dataset}\label{sec:real_world_dataset}

To further demonstrate the advantages of \textsc{CorDEL} over the DL baseline, we perform experiments on a real-world EL dataset, which casts more challenges compared to the public benchmark datasets. It is hard to classify a real-world dataset into one of the three types of EL task, since it is usually a mixture of them. In addition, a practical DL approach for EL needs to be stable, \textit{i.e.}, different training runs should lead to similar inference performance. This stability is crucial to make DL models reliable.

In order to evaluate the stability, we repeat each experiment for 10 times independently and report the mean and standard deviation over 10 runs. For the baseline \textsc{DeepMatcher}, we choose two versions, the simplest \textsc{DeepMatcher}-SIF and the most powerful \textsc{DeepMatcher}-Hybrid.

The comparisons between \textsc{CorDEL} and \textsc{DeepMatcher} are summarized in Table~\ref{table:real_world}. \textsc{CorDEL} has better and more stable performance in terms of both PRAUC and R@P=95$\%$. Formally, we conduct an unequal variance $t$-test on the PRAUC results between \textsc{CorDEL}-Attention and \textsc{DeepMatcher}-Hybrid. The $p$-value is 0.0069, indicating the improvement is statistically significant.

\begin{figure}
	\centering
	\includegraphics[width=0.4\textwidth]{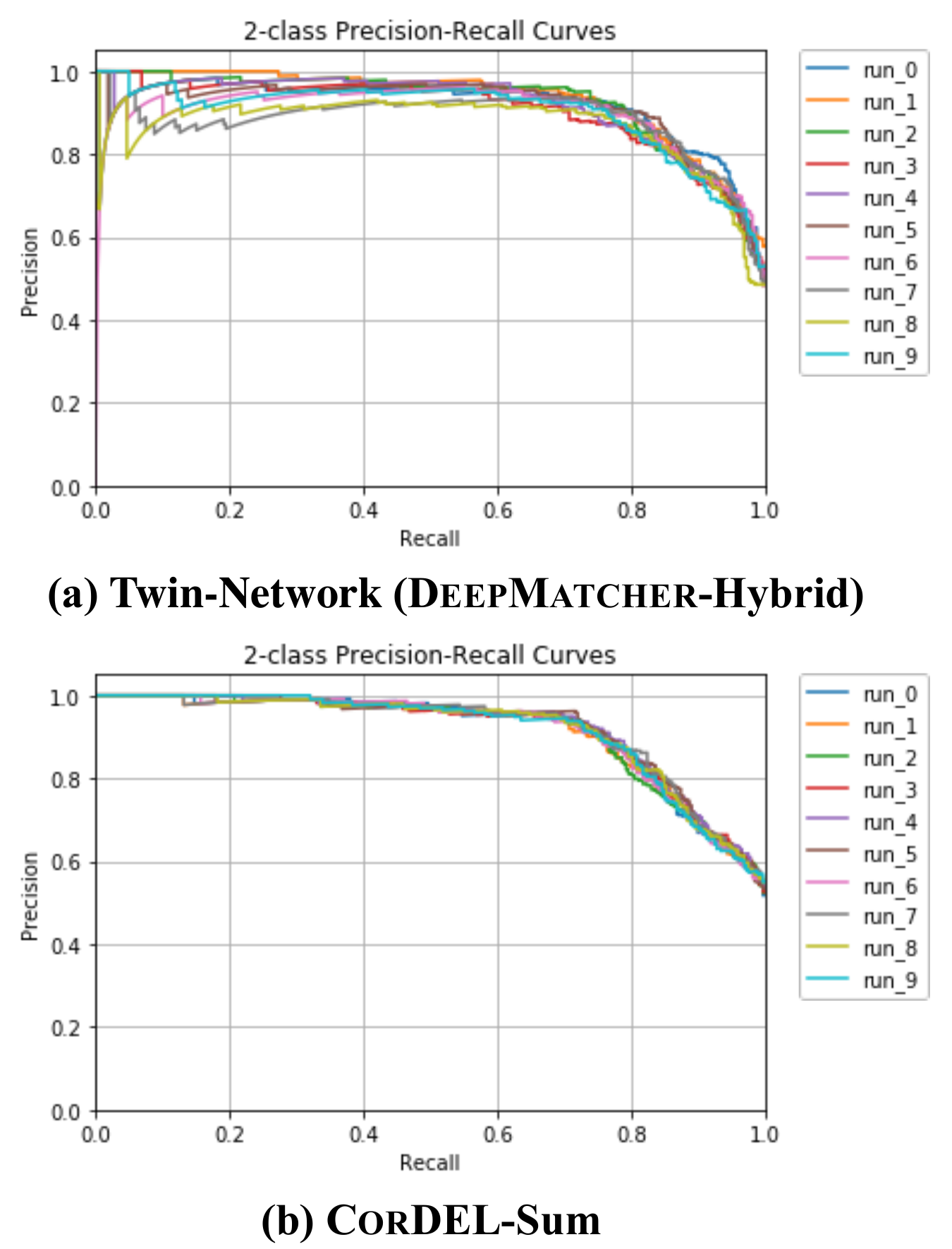}
	\caption{The precision-recall curves for \textsc{DeepMatcher}-Hybrid and \textsc{CorDEL}-Sum, with 10 independent runs for each of them. \textsc{CorDEL}-Sum is more stable with better performance especially for the high precision band.}
	\label{fig:plots}
\end{figure}

In order to show the superiority of \textsc{CorDEL} more directly, Figure~\ref{fig:plots} plots the precision-recall curves for \textsc{DeepMatcher}-Hybrid and \textsc{CorDEL}-Sum, with 10 independent runs for each of them. The instability of \textsc{DeepMatcher} can be easily observed. In addition, it is worth noting that, particularly, \textsc{CorDEL} has a much better and stable performance in the high-precision area.

\subsubsection{Efficiency Analysis}\label{sec:efficiency}

Another practical challenge in applying DL models on real-world EL tasks is the concern of efficiency. In particular, DL models tend to have a considerably large amount of training parameters, requiring large computational resources to train and deploy.

We compare the number of training parameters between \textsc{CorDEL} and \textsc{DeepMatcher} in the last column of Table~\ref{table:real_world}. We can see that even the simplest \textsc{DeepMatcher}-SIF has more parameters than \textsc{CorDEL}, while \textsc{CorDEL} yields much better performance as shown in the experiments above. In addition, the existing state-of-the-art DL approach, \textsc{DeepMatcher}-Hybrid, has millions of training parameters, preventing it from being applied on large-scale datasets. On the contrary, \textsc{CorDEL} is a light-weight and efficient DL approach.

\section{Conclusions}

In this work, we propose a novel contrastive DL approach for EL, called \textsc{CorDEL}. We point out the limitations of current twin-network DL models and motivate our work. We perform extensive experiments on both public benchmark datasets and a large real-world dataset for rigorous evaluations. The experimental results show the effectiveness of \textsc{CorDEL} with significant and consistent improvements in performance. Moreover, \textsc{CorDEL} is more efficient as a light-weight DL approach, and more reliable with stable performance.

\section*{Acknowledgment}

The authors would like to thank Christos Faloutsos, Andrew Borthwick, Yifan Ethan Xu, and Jialong Han for valuable suggestions..


\end{document}